\documentclass[pre,twocolumn,showpacs,preprintnumbers,amsmath,amssymb,floatfix]{revtex4}

\usepackage{graphicx}
\usepackage{dcolumn}
\usepackage{bm}

\usepackage{verbatim,vmargin,amsmath,calc, mathrsfs}
\usepackage{amsthm, amssymb, dsfont}
\usepackage{epsfig, pstricks, pst-plot,pst-node}
\usepackage{graphicx}
\usepackage{psfrag, wasysym}
\usepackage{multirow}

\newpsobject{showgrid}{psgrid}{subgriddiv=5}

\begin{document}
  
  \title{Evolving Networks through Deletion and Duplication}
  
  \author{Nadia Farid}
  \email{nadia.farid@imperial.ac.uk}
  \affiliation{Blackett Laboratory, Imperial College London,
    Prince Consort Road, London SW7 2BW, United Kingdom}

  \author{Kim Christensen}
  \email{k.christensen@imperial.ac.uk}
  \affiliation{Blackett Laboratory, Imperial College London,
    Prince Consort Road, London SW7 2BW, United Kingdom}
  
  \date{\today} 
  
  \begin{abstract}
    We introduce a minimalistic model based on dynamic node deletion
    and node duplication with heterodimerisation. The model is
    intended to capture the essential features of the evolution of
    protein interaction networks. We derive an exact two-step rate
    equation to describe the evolution of the degree distribution. We
    present results for the case of a fixed-size network. The results
    are based on the exact numerical solution to the rate equation
    which are consistent with Monte Carlo simulations of the model's
    dynamics. Power-law degree distributions with apparent exponents
    $<1$ were observed for generic parameter choices. However, a
    proper finite-size scaling analysis revealed that the actual
    critical exponent in such cases is equal to 1. We present a
    mean-field argument to determine the asymptotic value of the
    average degree, illustrating the existence of an attractive fixed
    point, and corroborate this result with numerical simulations of
    the first moment of the degree distribution as described by the
    two-step rate equation. Using the above results, we show that the
    apparent exponent is determined by the heterodimerisation
    probability. Our preliminary results are consistent with empirical
    data for a wide range of organisms, and we believe that through
    implementing some of the suggested modifications, the model could
    be well-suited to other types of biological and non-biological
    networks.
  \end{abstract}
  
  
  \maketitle

\section{Introduction}
In recent years, models of networks growing via single-node
duplication, divergence and mutation of links, considered in isolation
and combination, have assumed prominence in the literature on complex
networks. In a series of independent studies it was suggested that
these duplication-divergence-mutation models (hereafter called
`duplication models' for brevity) are good candidates to describe the
evolution and large-scale topological features of real protein-protein
interaction networks (PINs) in several organisms such as
\textit{S.Cerevisiae} and \textit{H.Pylori} \cite{WAGNER, CHUNG,
VAZQUEZ, KIM_KRAP, PS, SOLE_PS}.

In duplication models, proteins are represented by nodes, and a
pairwise interaction between any two proteins is represented by an
undirected link between the associated nodes, assumed to be fully
operative at all times. In a duplication event, a mother node is
chosen uniformly at random (u.a.r.) and each of its links are copied
to a newly created daughter node. Divergence refers to the subsequent
loss of links from the daughter node \cite{CHUNG, KIM_KRAP, PS,
SOLE_PS, RAVAL}, and/or the mother node \cite{ISPOLATOV}, or a shared
neighbour of both the daughter and mother node \cite{VAZQUEZ}. Often,
for simplicity, it is assumed that only the daughter node diverges. In
a mutation event, new links are added between the daughter node and
all other nodes in the network which are not already connected to the
mother node. Typically, one duplication-divergence event occurs at
each update step, and mutations, when considered, are modelled at a
rate much less than the divergence rate -- typically, one new link per
update step is added \cite{KIM_KRAP, PS, SOLE_PS}.

The idea of evolution through gene duplication is taken from biology
\cite{HUGHES, KENT, LYNCH, ZHANG}, where it was popularised in the
1970's by Ohno who conjectured that single and whole genome
duplications could provide the raw material for evolutionary
diversification \cite{OHNO}.  Whilst there is mounting evidence that
duplicate genes do occur in genomes \cite{WAGNER, FRIEDMAN, ZHANG,
GU_2, WOLFE, KELLIS}, it is widely acknowledged that little is known
about the details of the process of duplication itself, such as the
frequency of duplication events, the fate of duplicate genes, and the
frequency with which duplicate genes acquire novel functions
\cite{LYNCH, HUGHES, WAGNER, PRINCE}. In fact, whilst the microscopic
parameters are not yet known exactly, it has been suggested by Berg
\textit{et al.} \cite{BERG} that the rate of gain and loss of
interactions through mutations is at least an order of magnitude
higher than the growth rate due to duplications; as a result, the link
dynamics act as the dominant evolutionary force shaping the
large-scale statistical structure of the network, and not the gene
duplications; the slower gene duplication events only really affect
the size of the network.  Thus, it is still unclear if, and to what
extent, gene duplication is the dominant mechanism responsible for
the observed statistical features of PINs.  Moreover, gene deletion
and rearrangement are known to play important roles in the long-term
evolution of genomes \cite{KENT, KELLIS, AXELSEN} but have received
comparatively little attention in the literature on networks
addressing PINs \cite{NOORT, AXELSEN}.

In this paper, in a move to go beyond the duplication models and to
expand upon the emerging literature on network models of PINs
including gene deletion, we present a four-parameter model addressing
the scenario of network evolution through dynamic total node removal
in conjunction with growth by node duplication and
heterodimerisation. We refer to our network model as a
\textit{deletion-duplication-divergence-heterodimerisation} (DDDH)
model \cite{FOOTNOTE_1}. Although the model is primarily aimed at
describing the evolution of PINs, its description is kept general
enough so as to be applicable to a diverse range of complex systems
where components are added and removed throughout the system's
evolution.

In the wider literature on complex networks, previous studies which
have considered node deletion have generally regarded it as a
perturbation effect, used to test the tolerance of a network to random
and targeted attack \cite{BOLLOBAS, VAZQUEZ, JEONG, GU}.
More recently, the mechanism of dynamic node removal in conjunction
with growth by preferential attachment has been explored in
independent studies by Chung and Lu \cite{CHUNG_LU}, Cooper and Frieze
\cite{COOPER}, and Wang \cite{WANG}. They refer to such models as
\textit{growth-deletion} models. Since we consider growth by
duplication as opposed to growth by preferential attachment, our model
therefore also contributes to the literature on growth-deletion
models.

We focus our analysis on the degree distribution, $P(k)$,
characterising the probability for a node to have exactly $k$ links
\cite{BA, NEWMAN, DM}.
The degree distribution is the simplest topological feature to measure
and as a result, it has attracted and received the most attention in
the literature on complex networks.  It has been shown \cite{SOLE_PS,
KIM_KRAP, VAZQUEZ, ISPOLATOV, CHUNG, RAVAL} that the degree
distribution of networks generated by duplication models exhibits a
power-law tail, $P(k) \sim k^{-\gamma}$, for $k \gg 1$, where the
critical exponent $\gamma$ can be tuned such that it is in agreement
with observed exponents which are found to be in the range
$1<\gamma<3$ \cite{CHUNG, YOOK}.  Importantly, it has also been shown
that the degree distribution is a robust and generic property of PINs
common across different data sets -- an important consideration given
that current experimental techniques are notorious for suffering from
a high rate of false positives and false negatives \cite{BERG, YOOK}.


The paper is organised as follows. In Sec. II, we present the
formulation of the dynamics which describe the rules of evolution of
the network, defining the parameters and interpreting the rates. In
Sec. III, we present and discuss the exact two-step rate equation for
the evolution of the degree distribution. In Sec. IV, we present
results obtained from Monte Carlo (MC) simulations of the model and
the exact numerical solution of the rate equation for a generic choice
of parameter values. We discuss our results in Sec. V and end with a
conclusion in Sec. VI.

\section{DDDH Model}

We consider undirected networks where loops and multiple links are
forbidden. We start with a network of known size, $N$, and degree
distribution, $P(k,t=0)$, and allow it to evolve under the following
rules (see Fig. \ref{F:Toy_Model}):
\begin{figure}[h]
  \begin{pspicture}(-2,-0.5)(5,1)
    \rput(1.6,-3.2){\scalebox{0.6}{\includegraphics{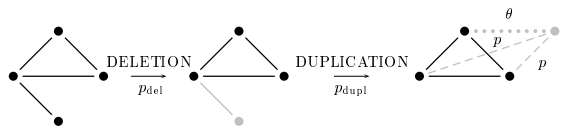}}}
  \end{pspicture}
  \caption{(Colour online) A schematic representation of a network
    evolving through deletion and duplication. (i) A node is chosen
    u.a.r. and deleted with probability $p_\textrm{del}$ (grey). (ii)
    A mother node is chosen u.a.r. and duplicated with probability
    $p_\textrm{dupl}$ (green $\rightarrow$ red). The links are
    retained with probability $p$ (dashed red line), and a further
    link is established between the daughter and mother node with
    probability $\theta$ (dotted blue line).}
  \label{F:Toy_Model}
\end{figure}

\noindent{{\textit{1.~Deletion.}} With probability $p_\textrm{del}$, a
node is chosen u.a.r. This node and all of its links are deleted
from the network.}\\\\
{{\textit{2.~Duplication-Divergence-Heterodimerisation.}}}  With
probability $p_\textrm{dupl}$, a mother node is chosen u.a.r. and
duplicated. This entails a new daughter node being added to the
network and linking to each of the neighbours of the mother node with
probability $p$. A further link is established between the daughter
and mother node with heterodimerisation probability $\theta$.\\

The evolution of the network is thus governed by four parameters:
$p_\textrm{del}, p_\textrm{dupl}, p$, $\theta$. In Sec. III, we cast
these rules into a concrete mathematical framework. First, however, it
is instructive to discuss the motivation behind the choice of dynamics
and its implications.

The mechanism of growth by duplication is preferred to the mechanism
of growth by preferential attachment in PIN network models as well as
many other network models since it reproduces the effects of
preferential attachment without having to artificially put the
mechanism in. In other words, it arises naturally from the dynamics:
nodes that have a large number of links are more likely to be
neighbours of a duplicating node, and hence are more likely to gain a
link to the newly created node. The DDDH model preserves this effect
and introduces another one: implicit preferential
\textit{detachment}. Nodes that have a large number of links are more
likely to be the neighbour of a node chosen for deletion, and
therefore will be more likely to lose a link each time a node is
deleted. It is interesting to note that these two effects do not
cancel out each other, as one might intuitively expect. The inclusion
of heterodimerisation, $\theta>0$, means that we do not consider
mutations in the traditional sense (as described above), but rather we
restrict the addition of new links to only occur between the mother
and the daughter node. Heterodimerisation is preferred over mutations
as it has been noted that the former increases the likelihood of
clique-formation -- a feature observed in real PINs -- whilst the
latter, in order to form the observed number of triads and higher
cliques, would require a prohibitively/physically unrealistic high
rate \cite{ISPOLATOV}. Moreover, in duplication models, whilst a lack
of random linking (mutations) has been found to destroy fine structure
such as the self-averaging and existence of a smooth degree
distribution \cite{KIM_KRAP}, the large-scale statistical features of
the final network do not depend on the existence of mutations
\cite{PS, KIM_KRAP, VAZQUEZ}.

From the general definition of the model's dynamics given above, one
can already determine specific features of the network, namely its
overall size, by focusing on particular choices of the model's
parameter values. For example, for a network that remains, on average,
fixed in size, $p_\textrm{del}=p_\textrm{dupl}$. We will refer to this
case as a `fixed-size' network. For a network that is monotonically
growing, on average, $p_\textrm{dupl}>p_\textrm{del}$; if
$p_\textrm{del}=0$ no nodes are removed from the network. Moreover, if
we fix $\theta=0$ and $0<p<1$ and $p_\textrm{dupl}=1$ for this case,
the DDDH model is equivalent to the duplication-divergence model
\cite{PS, SOLE_PS, ISPOLATOV, SOLE_FERNANDEZ}. For $p_\textrm{del}=0,
p_\textrm{dupl}=1, \theta=0$ and $p=1$ the duplication-divergence
model is equivalent to Polya's Urn \cite{CHUNG, ISPOLATOV, URNS}. For
an on average monotonically shrinking network,
$p_\textrm{dupl}<p_\textrm{del}$; if $p_\textrm{dupl}=0$ no new nodes
or links are added to the network \cite{FOOTNOTE_2}.  For a network
fluctuating in size the values of $p_\textrm{del}$ and
$p_\textrm{dupl}$ would be stochastic variables chosen anew at each
update step from a suitably chosen distribution \cite{FOOTNOTE_3}. If
$p=1$, the daughter node inherits all of the links; this is the case
of perfect, or full duplication. If $0<p<1$, the daughter node
inherits only some of the links; this represents imperfect, or partial
duplication. If $p=0$, the daughter node inherits none of the links,
that is, an isolated node is added to the network.

In this paper, we study the case of fixed-size networks,
$p_\textrm{del}=p_\textrm{dupl}=1$, with heterodimerisation $\theta>0$
and perfect duplication $p=1$, unless otherwise stated (in which case,
similar to Refs. \cite{CHUNG, KIM_KRAP, SOLE_PS} we assume that only
the daughter node diverges) \cite{FOOTNOTE_AIMS}.

In Sec. III, we apply the rate equation approach
\cite{KRAP_RE, KIM_KRAP} to study the evolution of the expected number
of nodes with $k$ links at time $t$, $f(k,t)=N_tP(k,t)$ where $N_t$ is
the total number of nodes at time $t$.

\section{DDDH model: Rate equation}
In this section, we present a two-step rate equation for the general
DDDH model, and then explain in detail the origin of each term.

The two-step rate equation for the DDDH model is given by,
\begin{widetext}
  \begin{subequations}\label{rateeqn}
    \allowdisplaybreaks
    \begin{align}
      f(k,t\!+\!1) =& f(k,t)-
      p_\textrm{del}\frac{f(k,t)}{N_t}-
      p_\textrm{del}\frac{k f(k,t)}{N_t}+
      p_\textrm{del}\frac{(k\!+\!1)f(k\!+\!1,t)}{N_t},\label{first}\\
      f(k,t\!+\!2)= &f(k,t\!+\!1)-p_\textrm{dupl}p\frac{k f(k,t\!+\!1)}{N_{t+1}}
      -p_\textrm{dupl}\theta \frac {f(k,t\!+\!1)}{N_{t+1}}\notag \\      
      \qquad &      +p_\textrm{dupl} p\frac{(k\!-\!1)f(k\!-\!1,t\!+\!1)}{N_{t+1}}
      + p_\textrm{dupl}\theta \frac{f(k\!-\!1,t\!+\!1)}{N_{t+1}} \notag\\ 
      \qquad &+ p_\textrm{dupl}\theta \sum_{j\geq(k\!-\!1)}
      \binom{j}{k\!-\!1}p^{(k-1)}(1\!-\!p)^{j-(k-1)}
      \frac{f(j,t\!+\!1)}{N_{t+1}}\notag\\
      \qquad &+ p_\textrm{dupl} (1\!-\!\theta) \sum_{j\!\geq k}
      \binom{j}{k}p^{k}(1\!-\!p)^{j-k}
      \frac{f(j,t\!+\!1)}{N_{t+1}}.\label{second}
    \end{align}
  \end{subequations}
\end{widetext}
Equation \eqref{rateeqn} is \textit{exact} and there are no
approximations in its derivation. It describes the DDDH process for
all parameters, $p_\textrm{del}, p_\textrm{dupl}, p, \theta$. It holds
for all $k \geq 0$, with $f(-1,t)=0$ and $f(k\!>\!k_\textrm{max},t)=0$
for all $t$. Moreover, $f(k,t)$ satisfies the following normalisation
conditions:
\begin{subequations}
\begin{align}
\sum_{k=0}^{\infty} f(k,t) &= N_t,\label{first} \\ \sum_{k=0}^{\infty}
k f(k,t) &= 2L_t,\label{second}
\end{align}
\end{subequations}
where $L_t$ is the total number of links at time $t$ in the
network. 

The distinct feature of the DDDH rate equation compared to rate
equations for duplication models is that it is defined in two
steps, and not one, reflecting the fact that we now include a node
deletion in addition to node duplication.
To keep the notation simple, we have written $t+1$ and $t+2$ in
Eq. \eqref{rateeqn} but one might have equally well written $t+1/2$
and $t+1$ to indicate that we \textit{only} observe the network after
\textit{both} the deletion and the duplication steps have been
completed.

Moreover, each of these actions is executed sequentially, highlighting
the fact that we clearly also make the distinction between this
process and the process of adding nodes at an `effective' duplication
rate, $p_\textrm{dupl}-p_\textrm{del}$, or merging nodes as is
Refs. \cite{SNEPPEN, DMS}, or substituting nodes where the duplication
of a node automatically implies the deletion of some other node as in
Ref. \cite{AXELSEN}. We note that the exact correspondence between
time, $t$, and real biological timescales is unclear, however, at this
stage.

In Eq. (1a) we consider the effects on the network of the deletion of
a node and the removal of its links. The probability a deletion event
occurs is given by $p_\textrm{del}$. The terms on the right-hand side
(RHS) are interpreted as follows. A \textit{loss} in the number of
nodes with degree $k$ at time $t+1$ from deletion will occur either if
the node deleted is of degree $k$ at time $t$, or if a neighbour (of
arbitrary degree) of a $k$-node is chosen for deletion, as the
$k$-node will lose a link and become a node of degree $k-1$. Since
every node has an equal probability of being deleted in a given time
step, a node of degree $k$ is chosen with probability $f(k,t)/N_t$
(second term); the probability that a neighbour of a $k$-node is
chosen for deletion is $kf(k,t)/N_t$ (third term). The final term on
the RHS represents a \textit{gain} in the number of nodes with degree
$k$ at time $t$. This can occur if the neighbour of a node with degree
$k+1$ is deleted. Given that a node of degree $k+1$ has $k+1$
neighbours, the probability a neighbour is chosen for deletion is
$(k+1)f(k+1,t)/N_t$.

In Eq. (1b) we consider the effects of node duplication and subsequent
heterodimerisation. The probability a duplication event occurs is
given by $p_\textrm{dupl}$, and the probability that subsequent
heterodimerisation occurs is given by $\theta$. All but the two final
terms on the RHS represent changes a duplication-heterodimerisation
event has on the existing nodes in the network; the last two terms
represent the daughter node's contribution.  

A \textit{loss} in the number of nodes with degree $k$ at time $t+2$
from duplication and heterodimerisation can occur in one of two ways.
If one of the neighbours (of arbitrary degree) of a node of degree $k$
at time $t+1$ is duplicated, the $k$-node will, with probability $p$
gain a link from duplication thus becoming a $k+1$ node at time $t+2$
(second term). Alternatively, if the mother node to be duplicated is
already of degree $k$ at time $t+1$, it will become a node with degree
$k+1$ at time $t+2$ by gaining a link to the daughter node via
heterodimerisation. The probability a node of degree $k$ is chosen for
duplication is given by $f(k,t+1)/N_{t+1}$, and the probability of
heterodimerisation is $\theta$ (third term).

We arrive at the fourth and fifth terms which describe the
\textit{gain} in the number of nodes with degree $k$ at time $t+2$ by
similar considerations. If one of the neighbours of a node of degree
$k-1$ at time $t+1$ is chosen for duplication, the $k-1$ node will, with
probability $p$, gain a link from duplication, thus becoming a $k$
node at time $t+2$ (fourth term).  Alternatively, if the mother node to
be duplicated is of degree $k-1$ at time $t+1$, it will become a node
with degree $k$ at time $t+2$ by gaining a link to the daughter node
via heterodimerisation (fifth term).

The final two terms on the RHS of Eq. (1b) account for the daughter
node's contribution. The sixth term is to account for the contribution
of the daughter node in the event it does establish a link, with
probability $\theta$, to the mother node. The seventh term is to
account for the case where it does not, which happens with probability
$(1-\theta)$.  Note the lower limits on these sums are not
identical. This is because if a link is established via
heterodimerisation, in order to become a node of degree $k$, the
daughter node is restricted to copying $k-1$ out of $j$ links, each
with probability $p$. However, if such a link is not established, the
daughter node is restricted to copying $k$ out of $j$ links of the
mother node.

Since the exact analytical solution to the rate equation in
Eq. \eqref{rateeqn} is not tractable at present, in Sec. IV we present
results obtained from the numerical solution of the exact
two-step rate equation and compare them with MC simulations of the
model. Unless otherwise stated, our analysis is based on the case of a
fixed-size network, $p_\textrm{del}=p_\textrm{dupl}=1$, evolving
through perfect duplication, $p=1$, with heterodimerisation,
$\theta>0$.


\section{DDDH model: Results}

\subsubsection{Comparison between exact numerical solution of the rate equation and MC simulations}

\begin{figure}[h]
  \begin{pspicture}(0,0)(7,6)
    \rput(3,3.0){\scalebox{0.325}
      {\rotatebox{270}{\includegraphics{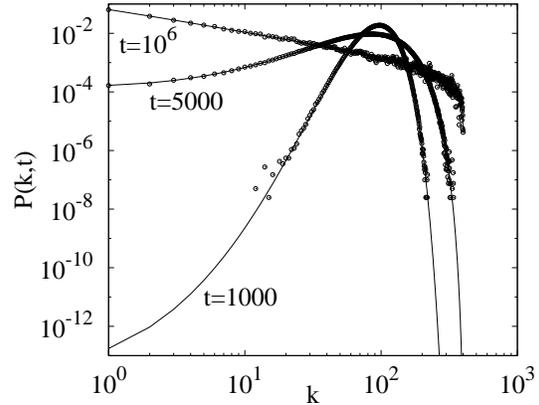}}}}
  \end{pspicture}
  \caption{Exact numerical (solid line) and MC (open circles) results
    of the degree distribution of a fixed-size network,
    $p_\textrm{del}\!=\!p_\textrm{dupl}\!=\!1$, with $N\!=\!400$
    nodes, evolving under perfect duplication, $p\!=\!1$, with
    heterodimerisation $\theta\!=\!0.1$.  Snapshots were taken at
    times $t\!=\!1000, 5000$ (transient regime) and $t=10^6$
    (stationary regime). The MC simulations are averaged over $10^5$
    realisations for $t\!=\!1000,5000$, and $3\times10^3$ realisations
    for $t\!=\!10^6$. The exact numerical results show excellent
    agreement with the MC simulations. The distribution in the
    stationary regime is well approximated by a power-law decay,
    $P(k)\sim k^{-\gamma}$ with an apparent exponent $\gamma\!=\!0.8$
    and sharp cut-off at $k_\textrm{cut-off}\!=\!399$. Note that the
    maximum degree a node can attain in networks of the type we
    consider is $N-1$.}
  \label{F:Trans_N400}
\end{figure}

Figure \ref{F:Trans_N400} displays the evolution of the degree
distribution obtained from the exact numerical solution of the rate
equation compared to MC simulations of the model. The stationary
regime is defined by $P(k,t)=P(k,t+2)$.

We start with an initial network of $N=400$ nodes, with a random
degree distribution centred around $k_\textrm{init}=100$, and iterate
the rules with the following parameter settings:
$p_\textrm{del}=p_\textrm{dupl}=1$ (fixed-size network), $p=1$
(perfect duplication), with heterodimerisation $\theta=0.1$. The value
for $\theta$ was chosen as such as it is believed that
heterodimerisation occurs at a rate not greater than $0.1$
\cite{ISPOLATOV}. In Fig. \ref{F:Trans_N400}, we show two snapshots
of the network in the transient regime when $t=1000, 5000$,
respectively, and one in the stationary regime for $t=10^6$. The MC
simulations are averaged over $10^5$ realisations for $t=1000, 5000$ and
$3\times10^3$ realisations for $t=10^6$.

There is excellent agreement between the exact numerical solution and
MC results, lending support to our statement earlier that there are no
approximations involved in our derivation of the rate equation. We
have verified through extensive simulations that the agreement holds
true over a range of $p_\textrm{del}, p_\textrm{dupl}, p$, and
$\theta$ values. Hence, all remaining figures are generated using data
obtained from the exact numerical solution of the rate equation only,
hereafter referred to as `exact numerical results'. The interesting
feature to note is that even for the simplest realisation of the DDDH
model which we have presented in Fig.  \ref{F:Trans_N400}, fat-tailed
degree distributions are obtained in the stationary regime ($t=10^6$
curve). This is in stark contrast to the duplication models where only
the case of imperfect duplication leads to a power law \cite{CHUNG}.
In the following section, we describe quantitatively the exact form of
the stationary degree distribution.

\subsubsection{Scaling for $p_\textrm{del}=p_\textrm{dupl}=p=1, 0<\theta \leq 1/2$}

We are interested in quantifying the form of the degree distribution,
in the stationary regime, as a function of the model's
parameters. Given that we are, for the moment, investigating
fixed-size networks evolving under perfect duplication,
$p_\textrm{del}, p_\textrm{dupl}$ and $p$ are all fixed. This reduces
the number of variables to just one: $\theta$.  However, given that we
are observing the degree distribution for specific fixed network
sizes, we have $N$ as another variable in the problem. Hence, we would
like to know how $P(k)$ depends on $\theta$ and $N$. In order to
investigate this, we have performed numerical simulations for the
following two cases: (i) Fixed $\theta$, varying $N$, and (ii) Fixed
$N$, varying $\theta$. We discuss (i) in this subsection, and (ii) in
Sec. IV. 3.

\begin{figure}[h] 
  \begin{pspicture}(0,0.475)(7,11.0)
    \rput(3.4,8.5){\scalebox{0.3}
      {\rotatebox{270}{\includegraphics{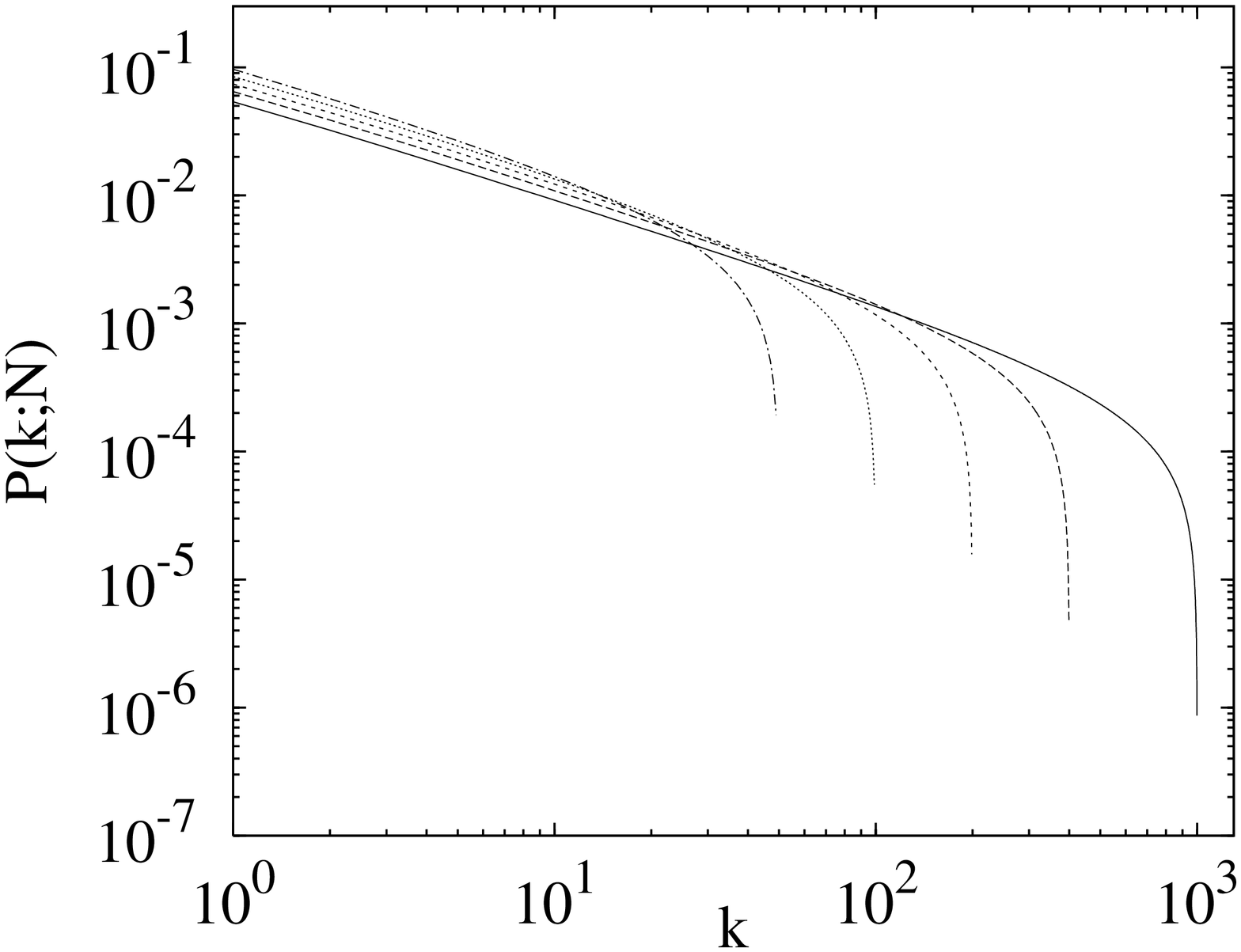}}}}
    \rput(3.4,3.1){\scalebox{0.3}
      {\rotatebox{270}{\includegraphics{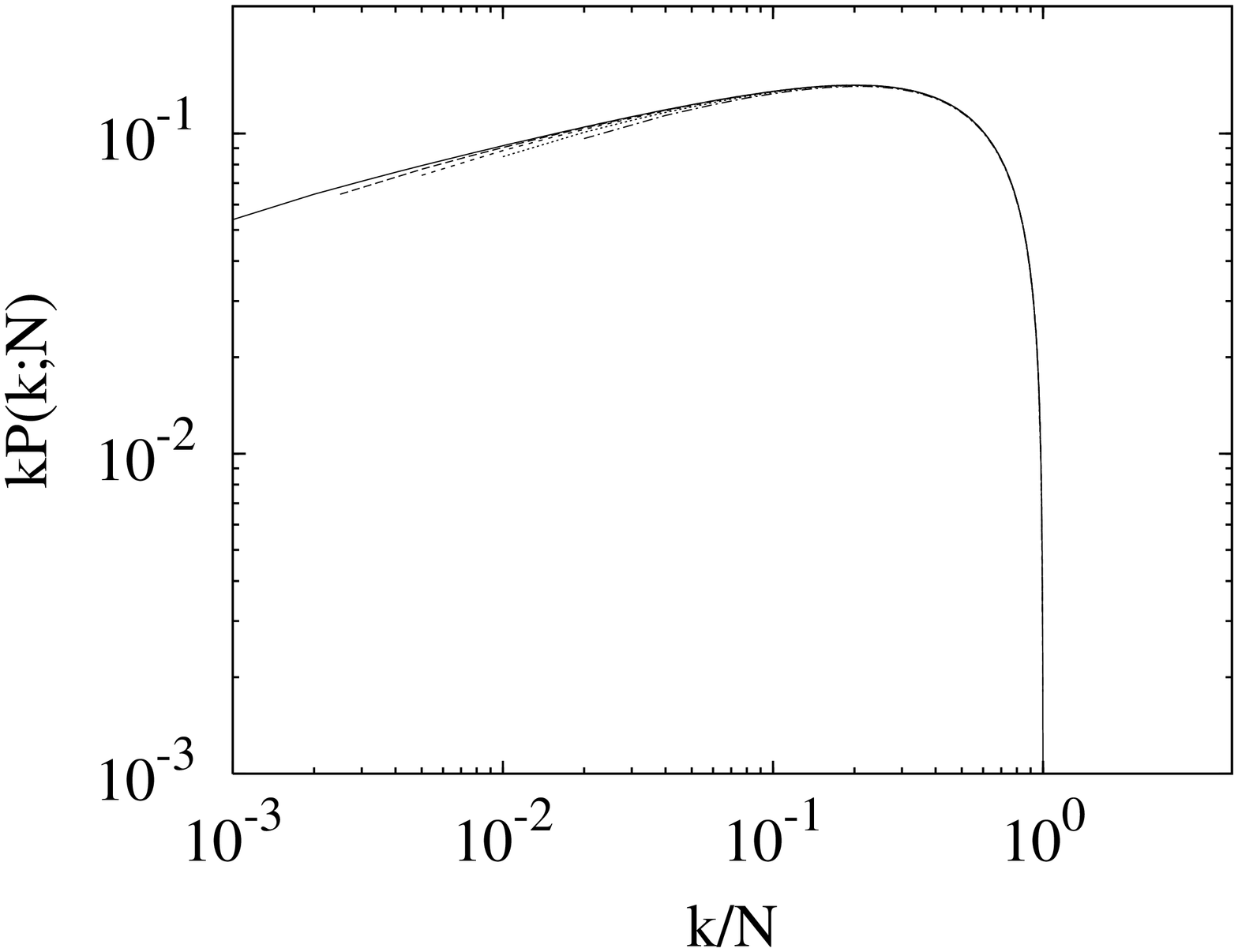}}}}
    \rput(6.25,10.5){(a)}
    \rput(6.25,5.1){(b)}
  \end{pspicture}
  \caption{(a) Exact numerical results of the degree distribution in
    the stationary regime for fixed-size networks,
    $p_\textrm{del}\!=\!p_\textrm{dupl}\!=\!1$, where $N\!=\!50, 100,
    200, 400, 1000$ (marked with lines of increasing dash-length) each
    having evolved through perfect duplication $p\!=\!1$ with
    heterodimerisation $\theta\!=\!0.1$. There is no typical
    node-degree. For large node-degrees, the degree distribution is
    well-approximated by a power-law decay, $P(k;N)\sim k^{-\gamma}$,
    with an apparent exponent $\gamma=0.8$. The power law is
    characterised by a sharp cut-off at $k_\textrm{max}\!=\!N-1$,
    which increases with increasing system size. Note that the maximum
    degree a node can attain in networks of the type we consider is
    $N-1$. (b) Data collapse of the exact numerical results of the
    degree distribution is obtained by plotting the transformed
    probability density $kP(k;N)$ vs. the rescaled degree $k/N$. The
    curves collapse onto the graph of the scaling function,
    $\widetilde{\mathcal{G}}(k/N)\!=\!\frac{1}{\Gamma(1-\gamma)\Gamma(1+\gamma)}(k/N)^{1-\gamma}(1-k/N)^{\gamma}$,
    see Eq. \eqref{eqn:k1}.}
  \label{Fig:DataCollapse}
\end{figure}

Figure \ref{Fig:DataCollapse}(a) shows the exact numerical results of
the degree distribution, $P(k;N)$ versus $k$ on a double logarithmic
plot in the stationary regime for $\theta=0.1$ and networks of
increasing $N$, specifically, $N={50, 100, 200, 400, 1000}$
nodes. There are clear power-law fluctuations in the node degrees
present in the network, implying an appreciable probability of finding
a node with degree, $1 \leq k \ll k_\textrm{max}$ in the
network. $k_\textrm{max}$ marks the cross-over between a power-law
decay and a rapid decay in $P(k;N)$. In particular, $k_\textrm{max}$
represents a characteristic scale in the node degree resulting from
the finite size of the networks we can study numerically. Hence, we
can say that $P(k;N)$ decays as a power law for $1\leq k\ll k_\textrm{max}$
and has a sharp cut-off for $k \gg k_\textrm{max}$, which can be expressed
informally as,
\begin{eqnarray}\label{eqn:general_PL}
  P(k;N) \propto \left \{ 
  \begin{array}{ll} 
    k^{-\gamma}              & \mbox{$1\leq k\ll k_\textrm{max}$}\\ 
    \textrm{sharp cut-off, } & \mbox{$k \gg k_\textrm{max}.$} 
  \end{array} \right.
\end{eqnarray}
From simulations, we find that $k_\textrm{max}=N-1$ which is
equivalent to the maximum possible degree that a node in the network
can acquire. This implies that $k_\textrm{max}$ increases linearly
with increasing network size, $N$, hence, in the limit of $N
\rightarrow \infty$, the characteristic scale diverges and a pure
power law is recovered, as expected. In the region,
$1\leq k\ll k_\textrm{max}$, we find that the gradient of the lines in
Fig. \ref{Fig:DataCollapse}(a) are well-approximated by an `apparent'
exponent, $\gamma=0.8$ and we will shortly demonstrate that
$\gamma=1-2\theta$. Generally speaking, an exponent less than $1$ is
unusual, and seems to contradict certain known results about scaling
functions. However, we will be able to resolve this apparent
contradiction in Sec. IV. 3.

With the above discussion in mind, we propose the following general
ansatz for $P(k;N)$,
\begin{align}\label{eqn:scaling}
P(k;N) &= a(N) k^{-\gamma} \mathcal{G}(k/N),
\end{align}
and, assuming for simplicity that $k_\textrm{max}$ is approximated by
$N$, the equation is valid for $1 \leq k \leq N$, where $N \gg 1$ and
$\gamma<1$. In Eq. \eqref{eqn:scaling}, $a(N)$ is a prefactor,
dependent on the network size, and $\mathcal{G}(x)$ is the cut-off
function, dependent on the rescaled variable $k/N$. $\mathcal{G}(x)$
is required to fall-off fast enough to ensure $P(k;N)$ is finite and
integrable. From Fig. \ref{Fig:DataCollapse}(a) it seems reasonable to
assume that $\mathcal{G}(x)$ is constant for $x\ll 1$ and decays
abruptly for $x\gg 1$.


Note that Eq. \eqref{eqn:scaling} is \textit{not} a finite-size
scaling (FSS) ansatz since the prefactor, rather than being a constant
(as is typical), is $N$-dependent. This difference turns out to be
important as it leads to an interesting result about the critical
exponent which we will demonstrate below.

In theory, we can use the fact that the probability density function
must be properly normalised,
\begin{align}\label{eqn:norm}
  \int_{1}^{\infty} P(k;N) dk \equiv 1,
\end{align}
to derive an expression for $a(N)$. However, without knowing \textit{a
priori} the correct scaling for $P(k;N)$, we can only guess at the
form of $\mathcal{G}(x)$. For simplicity, we assume that the cut-off function,
$\mathcal{G}(x)$, is of the form,
\begin{eqnarray}\label{eqn:scaling_function}
  \mathcal{G}(x) = \left \{ 
  \begin{array}{ll} 
    (1- k/N)^{\gamma}  &\mbox{for $1\leq k\leq N$}\\ 
    0 &\mbox{otherwise}  \end{array}\right.
\end{eqnarray}
(which is in-keeping with our stated requirements for the form of
$\mathcal{G}(x)$ and is an excellent fit to the
numerics). Substituting this into Eq. \eqref{eqn:scaling} we find that
normalisation requires,
\begin{align}\label{eqn:scaling_function_sub}
  \int_{1}^{N} a(N) k^{-\gamma} (1-k/N)^{\gamma} dk \equiv 1.
\end{align}
Evaluating the LHS of Eq. \eqref{eqn:scaling_function_sub} we find for
$N\gg 1$,
\begin{align}
  a(N)N^{1-\gamma}\Gamma(1-\gamma)\Gamma(1+\gamma) = 1,
\end{align}
and it immediately follows that,
\begin{align}\label{eqn:aN}
  a(N) &= \frac{N^{\gamma-1}}{\Gamma(1-\gamma)\Gamma(1+\gamma)}.
\end{align}                                       
Using this result for $a(N)$, we can recast Eq. \eqref{eqn:scaling}
into a scaling ansatz such that an \textit{actual} critical exponent
equal to $1$ is obtained, as follows,
\begin{align}\label{eqn:k1}
  P(k;N) &= \frac{N^{\gamma-1}}{\Gamma(1\!-\!\gamma)\Gamma(1\!+\!\gamma)}k^{-\gamma} \mathcal{G}(k/N)\nonumber\\
  &= k^{-1}\frac{1}{\Gamma(1\!-\!\gamma)\Gamma(1\!+\!\gamma)} \frac{k^{1-\gamma}}{N^{1-\gamma}}\mathcal{G}(k/N)\nonumber\\
  &= k^{-1}\widetilde{\mathcal{G}}(k/N),
\end{align}
where $\widetilde{\mathcal{G}}(x) =
\frac{1}{\Gamma(1\!-\!\gamma)\Gamma(1\!+\!\gamma)}
x^{1-\gamma}\mathcal{G}(x)$.  Equation \eqref{eqn:k1} is our `proper'
FSS ansatz. We have shown using consistent arguments that
Eq. \eqref{eqn:scaling} can be recast into Eq. \eqref{eqn:k1} assuming
the cut-off function, $\mathcal{G}(x)$ is of the form given in
Eq. \eqref{eqn:scaling_function}, and using the requirement that the
probability density function must be properly normalised to derive an
expression for the $N$-dependent prefactor, $a(N)$. The point of
interest is that on the LHS of Eq. \eqref{eqn:k1} the leading
power-law term has attained a fixed value equal to $1$, independent of
$\gamma$. Thus, even if the apparent measured exponent, $\gamma$, is
in the range $[0,1)$ the actual critical exponent is always equal to
1.

To test the validity of the scaling ansatz given in
Eq.\eqref{eqn:k1}, we have, in Fig. \ref{Fig:DataCollapse}(b),
plotted the transformed probability density $kP(k;N)$ versus
the rescaled variable $k/N$ using the same data as in
Fig. \ref{Fig:DataCollapse}(a). Multiplying both sides of
Eq. \eqref{eqn:k1} by $k$ we get,
\begin{align}\label{eqn:DataCollapse}
  kP(k;N) = \widetilde{\mathcal{G}}(k/N).
\end{align}
Therefore, we expect the curves to collapse onto the curve
$\widetilde{\mathcal{G}}(k/N)=\frac{1}{\Gamma(1\!-\!\gamma)\Gamma(1\!+\!\gamma)}(k/N)^{1-\gamma}(1-k/N)^{\gamma}$,
with the gradient of the slope to be equal to $1-\gamma$. As shown in
Fig. \ref{Fig:DataCollapse}(b), a convincing data collapse is
obtained, with all curves collapsing onto the scaling function
described by $\widetilde{\mathcal{G}}(x)$ with
$\gamma=1-2\theta=0.8$. We have repeated the data collapse for
different values of $\theta$, and observed a convincing data collapse
in all cases, with all curves collapsing onto the scaling function
described by, $\widetilde{\mathcal{G}}(x)$ with $\gamma=1-2\theta$.

Together, Eq. \eqref{eqn:scaling} and the success of the data collapse
in Fig. \ref{Fig:DataCollapse}(b) demonstrate that the degree
distribution for any network size, $N$, is determined by the scaling
function $\widetilde{\mathcal{G}}(x)$. This means that we can deduce
the degree distribution for any network size, $N$, without having to
actually perform the numerical simulation itself. Hence, our results
are applicable to networks larger than those which we have
demonstrated directly, that is for $N>10^3$. This is in contrast to
the duplication models considered in \cite{ISPOLATOV} where the
networks generated do not attain power-law degree distributions even
for very large networks.

Thus far, we have not yet justified the relation given between the
apparent exponent and the parameter $\theta$. In the following
section, we derive a result for the average degree, $\langle
k\rangle$. We then demonstrate how we can use this result to find an
expression for the apparent exponent $\gamma$ in terms of the
parameter, $\theta$.

\subsubsection{Mean-field Equation for the Average Degree}

The average degree, $\langle k \rangle$, can be determined in various
different ways. Ideally, one would be able to calculate it directly
from the two-step rate equation for the evolution of the degree
distribution in Eq. \eqref{rateeqn}. This would be achieved by taking
the first moment of the normalised degree distribution, $P(k,t) =
f(k,t)/N_t$, according to,
\begin{equation}\label{firstmoment}
  \langle k \rangle_{t} = \int_{1}^{\infty} k P(k,t) dk.
\end{equation}
Strictly speaking, where we write integration signs we should have
sums, as we are dealing with a discrete probability distribution.
Either way, the solution we are interested in is $lim_{t\rightarrow
\infty} \langle k\rangle_{t} = \langle k\rangle$, which is
analytically intractable. In order to get around this problem we have
calculated this quantity numerically, and compared this result to the
asymptotic value of $\langle k \rangle$ obtained as a solution to a
mean-field rate equation approach (described below).

\begin{figure}[h]
  \begin{pspicture}(0,0)(7,5)
    \rput(3.5,2.5){\scalebox{0.3}{\rotatebox{270}{\includegraphics{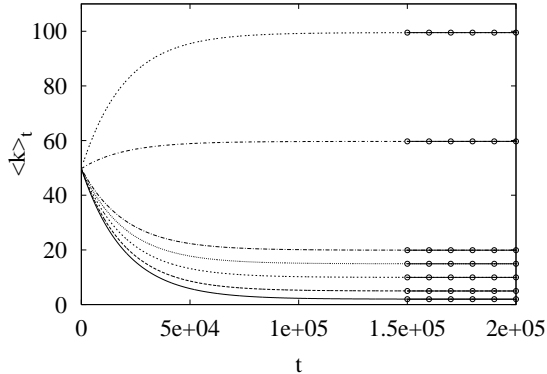}}}}
  \end{pspicture}
  \caption{Exact numerical results for the average degree, $\langle
    k\rangle$, versus time for a fixed-size network,
    $p_\textrm{del}\!=\!p_\textrm{dupl}\!=\!1$, with $N\!=\!200$ nodes,
    and $k_\textrm{initial}\!=\!50$. The network evolved through perfect
    duplication and increasing $\theta\!=\!0.01, 0.025, 0.05, 0.075,
    0.1, 0.3, 0.5$ (marked with lines of decreasing dash-length). In
    each case, $\langle k \rangle$ reaches a stationary value whose
    value is identical to that predicted analytically (circles),
    $\langle k\rangle_\textrm{MF}=\theta(N-1)$, see Eq. \eqref{eqn:MEAN_FIELD}.}
  \label{F:AveK}
\end{figure}

Figure \ref{F:AveK} illustrates the exact numerical results for the
time-evolution of $\langle k \rangle$ for $\theta \in [0.01, 0.5]$ and
clearly illustrates the existence of a stationary asymptotic $\langle
k \rangle$.  We now describe our mean-field argument to determine the
asymptotic value of the average degree, $\langle k \rangle$. At each
time step, for each node we delete we lose on average, $\langle
k\rangle_{t}$ links, and for each node we duplicate we gain on
average, $p\langle k\rangle_{t+1} + \theta$ links. Therefore, the net
change in the number of links, $\Delta L = -\langle k\rangle_{t} +
p\langle k\rangle_{t+1} + \theta$. For the case of $p=1$, we can
rewrite this as, $\Delta L = -\frac{2L_t}{N_t} +
\frac{2L_{t+1}}{N_{t+1}} + \theta$, where we have used the standard
relation, $\langle k\rangle=2L/N$. Imposing the condition $\Delta
L=0$, which is valid in the stationary regime, $t \rightarrow \infty$,
we find,
\begin{equation}\label{eqn:MEAN_FIELD}
  \langle k \rangle_\textrm{MF} = \theta(N-1)
\end{equation}
for a fixed-size network evolving under perfect duplication for
arbitrary $\theta$. So, for a network of size $N=200$, for example, we
predict, using Eq. \eqref{eqn:MEAN_FIELD}, $\langle k \rangle=1.99,
19.9, 99.5$ for $\theta=0.01, 0.1, 0.5$ respectively. We can compare
this prediction with the exact numerical results for the first moment
of the degree distribution. As shown in Fig.\ref{F:AveK}, there is
exact agreement between the asymptotic value of the average
degree determined from the mean-field calculation and the exact
numerical results (for $t>10^5$).
Thus, both the mean-field calculation and the exact numerical results,
as illustrated in Fig. \ref{F:AveK}, demonstrate the existence of an
attractive fixed point in the average degree, $\langle k\rangle$. This
is clearly related to the existence of a stationary degree
distribution, that is, $P(k,t) = P(k,t+2)$.

Equation \eqref{eqn:MEAN_FIELD} highlights the importance of
accounting for the mechanism of heterodimerisation for finite-sized
networks, as it shows that the network self-organises to a stationary
state where the average degree $\langle k \rangle$ is constant, and
determined by the system size, $N$ \textit{and} the probability for
heterodimerisation, $\theta$. Since real PINs are of finite size,
typically with no more than $10^4$ nodes (see Table I), the point of
information regarding the role of $\theta$ in finite-sized networks is
of significance.

We have also verified through further simulations varying $p$ such
that $p<1$ for constant $\theta$ and $N$ clearly dramatically reduces
$\langle k\rangle$, as one would expect, although the precise nature
of this effect has not yet been quantified and seems to be
non-trivial. Thus, Eq. \eqref{eqn:MEAN_FIELD} actually gives an upper
bound for $\langle k\rangle$.

An alternative method for calculating $\langle k \rangle$
analytically, is to calculate the first moment of the degree
distribution as expressed in Eq. \eqref{eqn:k1}, $\langle k
\rangle_\textrm{SF}$. This turns out to be very useful as far as
determining an expression for the apparent exponent, $\gamma$, in
terms of $\theta$.  We find that,
\begin{align} \label{eqn:FirstMoment_ScalingFunction}
  \langle k \rangle_\textrm{SF} &= \int_{1}^{N} kP(k;N) dk \nonumber\\
  &= \int_{1}^{N} \frac{1}{\Gamma(1\!-\!\gamma)\Gamma(1\!+\!\gamma)} \left(\frac{k}{N}\right)^{1\!-\!\gamma}\left(1\!-\!\frac{k}{N}\right)^{\gamma} dk\nonumber\\
  &= \frac{N}{2} \frac{\Gamma(2-\gamma)}{\Gamma(1-\gamma)} \quad\mbox{for $N\rightarrow \infty$}\nonumber\\
  &= \frac{N}{2} (1-\gamma). 
\end{align}
Since we already know that ${\langle k\rangle}_\textrm{MF} =
\theta(N-1)$ from Eq. \eqref{eqn:MEAN_FIELD}, the RHS of
Eq. \eqref{eqn:FirstMoment_ScalingFunction} must be equivalent to
Eq. \eqref{eqn:MEAN_FIELD}, hence,
\begin{align}
  \gamma = 1- 2\theta.
\end{align}
This justifies our previous finding in Sec. IV. 2 that the apparent
exponent is $\gamma=1-2\theta$.  In the following section, we
investigate the effect on the degree distribution of varying
$\theta \in (0,0.5)$ for fixed $N$, completing the analysis of the two
scenarios outlined at the beginning of Sec. IV. 2.


\subsubsection{Effects of varying $\theta$}

\begin{figure}[h]
  \begin{pspicture}(0,0)(7,5)
    \rput(3,2.5){\scalebox{0.3}
      {\rotatebox{270}{\includegraphics{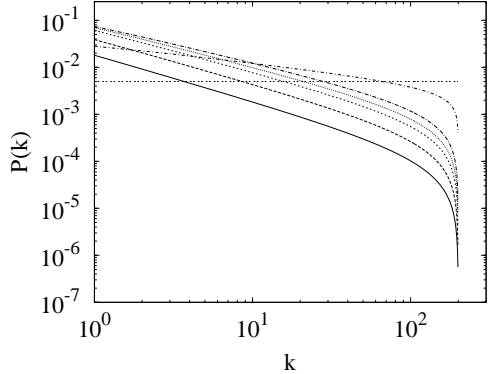}}}}
  \end{pspicture}
  \caption{Exact numerical results for the degree distribution in the
    stationary regime for a fixed-size network,
    $p_\textrm{del}\!=\!p_\textrm{dupl}\!=\!1$, with $N\!=\!200$ nodes,
    and $k_\textrm{initial}\!=\!50$. The network evolved through perfect
    duplication and increasing $\theta\!=\!0.01, 0.025, 0.05, 0.07, 0.1,
    0.3, 0.5$ (marked with lines of decreasing dash-length). The value
    $\theta\!=\!0.5$ marks a change in behaviour from a power-law decay
    with a negative exponent to a uniform distribution.}
  \label{F:VaryTheta}
\end{figure}

Figure \ref{F:VaryTheta} illustrates the topological effect of varying
the probability to heterodimerise in the range $0<\theta\leq 0.5$, in
a fixed-size network ($N=200$), evolving through perfect duplication.

We find that the degree distribution exhibits a power law with a slope
that varies with $\theta$ according to $\gamma=1-2\theta$, reaching a
uniform distribution at a value of $\theta=0.5$. We have repeated the
above simulations for a range of network sizes, up to $N=1000$, and
confirmed this behaviour. Hence, the result for $\gamma$ is consistent
with our previous findings in Sec. IV. 2.

Current estimates from empirical data for yeast, fly and human PINs
indicate that $\theta$ never exceeds $0.1$ \cite{ISPOLATOV}. Since we
observe power-law degree distributions in the range $0<\theta<0.5$, a
value of $\theta<0.1$ in our model is consistent with empirical data.
The fact that $\gamma=1-2\theta$ might go some way towards explaining
why in the duplication model considered in Ref. \cite{ISPOLATOV}, for
realistic values of $\theta<0.1$ their results were not affected.

\subsubsection{Effects of varying $p$}

Up until now, we have been investigating the effects of varying $\theta$
and $N$, for fixed $p_\textrm{del}=p_\textrm{dupl}=p=1$, on the degree
distribution. We now report our findings for a third possible
scenario: the effect of varying $p$, for fixed $\theta$ and $N$
(keeping $p_\textrm{del}=p_\textrm{dupl}=1$, as before).

\begin{figure}[h]
\begin{pspicture}(0,0)(7,5)
  \rput(3,2.5){\scalebox{0.3}{\rotatebox{270}{\includegraphics{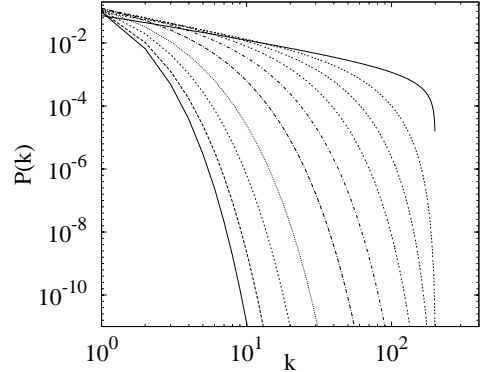}}}}
\end{pspicture}
\caption{Exact numerical results for the degree distribution in the
stationary regime for a fixed-size network,
$p_\textrm{del}\!=\!p_\textrm{dupl}\!=\!1$, with $N\!=\!200$ nodes,
for increasing duplication probabilities $p\!=\!0.1, 0.2, 0.4, 0.6,
0.8, 0.9, 0.95, 0.975, 0.99, 1$ (marked with lines of decreasing
dash-length) and heterodimerisation $\theta\!=\!0.1$. There is a
marked difference between the curves generated with $p\!<\!1$ and with
$p\!=\!1$. Whereas the latter is described by a power-law decay, the
former are not and a characteristic degree is present.}
\label{F:Vary_p_static}
\end{figure}

Figure \ref{F:Vary_p_static} illustrates the topological effect of
varying $p$ in a fixed-size network with heterodimerisation
$\theta=0.1$. There is clearly a marked difference between the curve
for $p=1$ and the family of curves for $p<1$. We established in
Sec. IV. 2-4 that for $p=1$ and $\theta \in (0,0.5)$, the degree
distribution is well approximated by a power-law decay. We now see
that for $p<1$, the power-law behaviour is no longer observed and a
characteristic degree is present. We have confirmed this result for a
range of network sizes, specifically, $N=50,100,400,1000$. Moreover,
we have found that the second moment $\langle k^{2} \rangle$ does not
diverge with increasing network size as one would expect if, in the
limit of $N \rightarrow \infty$, the degree distribution were indeed
described by a power-law decay.

We can offer a simple heuristic argument to account for the difference
between the case $p=1$ and $p<1$. We believe that it is directly
related to the choices we have made for the remaining parameters of
the model, namely $p_\textrm{del}=p_\textrm{dupl}=1$ and $\theta>0$.
For $p<1$ at each duplication event only some of the links of the
mother node are copied by the daughter node, whereas for $p=1$, all of
the mother node's links are copied. Given that $p_\textrm{del}=1$ we
delete a node and all of its links at each time step, and thus at each
duplication event, if $p<1$, we do not compensate for the loss of
links incurred through the deletion process. Hence, the repeated
application of duplication events with $p<1$, given that
$p_\textrm{del}=1$ accounts for the fact that the observed degree
distribution does not follow a power-law decay but rather has a
characteristic degree present.

\subsubsection{Comment on relation to biological data}

We can test the suitability of the DDDH model as a representation of
the evolution of PINs by comparing our results against empirical data.
Using data cited in Ref. \cite{ISPOLATOV} we can obtain values of the
number of proteins in the network and estimates of the average degree
of the PINs for yeast, fly, and human. From these, we can derive
estimates for $\theta$ using Eq. \eqref{eqn:MEAN_FIELD} and the
apparent scaling exponent, $\gamma=1-2\theta$. The results are
tabulated in Table I.  
\newcommand\T{\rule{0pt}{0.175in}}
\begin{table}[h]
  \begin{tabular}{|p{0.7in}|p{0.45in}|p{0.4in}|p{0.6in}|p{0.3in}|}
    \hline
    Data \T      & N      &$\langle k\rangle$ &$\theta$   &$\gamma$\\
    \hline
    Yeast (I)   & $4873$ & $6.6$             & $0.0014$  & 1.0\\
    \hline
    Yeast (II) & $5397$ & $29.2$            & $0.0054$  & 1.0\\ 
    \hline
    Fly        & $6954$ & $5.9$             & $0.00085$ & 1.0\\
    \hline
    Human      & $5275$ & $5.7$             & $0.0011$  & 1.0\\
    \hline
  \end{tabular}
  \caption{Comparison between empirical data \cite{ISPOLATOV,
      VON_MERING} and the fixed-size, perfect duplication DDDH model.
      Values of $\theta$ and $\gamma$ are quoted to 2 decimal places.}
\end{table}

For all data sets, the calculated value of $\theta$ is of the order of
$10^{-3}$ which agrees well with the fact that it is believed
heterodimerisation occurs at a rate not greater than $0.1$
\cite{ISPOLATOV}. Moreover, the corresponding apparent scaling
exponent extracted in all cases is found to be $\gamma=1.0$. This is
consistent with our FSS analysis and corroborates
Ref. \cite{ISPOLATOV} where a power law with degree exponent $\gamma
\approx 1.1$ is given for the data for yeast derived from
\cite{VON_MERING}.

\section{Discussion}

The results in Sec. IV indicate that the degree distribution of
fixed-size networks where the rate of node deletion is equal to the
rate of node duplication, $p_\textrm{del}=p_\textrm{dupl}=1$, is
dependent on several features. We summarise these as follows. (a) In
order to observe a power-law degree distribution it is necessary to
have $p=1$, that is, perfect duplication, and $0<\theta<0.5$. (b) No
power-law is observed in the degree distribution for $p<1$, or for
$p=1$ and $\theta \geq0.5$. (c) In all cases, it is necessary for
$\theta$ to be non-zero in order for a positive average degree to be
obtained in the stationary regime since for $\theta=0, \langle k
\rangle=0$. (d) In cases where a positive average degree is obtained,
it is notable that the network self-organises into a stationary
state. We see this as an advantageous feature of this model and
comment that this is in contrast to some other network models, such as
in Ref. \cite{BERG}, where the average degree is a fixed parameter in
the model, or the duplication models where the average degree scales
with the network size. (e) Our FSS analysis indicates that for fixed
$\theta \in (0,0.5)$, the scaling exponent of the
associated (proper) FSS ansatz is fixed and equal to 1. The new result
is obtained through the inclusion of a system-size dependency in the prefactor, $a(N)$,
necessary when the apparent exponent $\gamma<1$, which results in the scaling function being
recast in such a way that the only relevant scale parameterising the
system is determined by the cut-off, given by $N$ in our case. This
example illustrates that it is necessary to be cautious when doing a
FSS ansatz in systems where the apparent exponent appears to be less
than 1 \cite{MATT}. (f) Estimates of $\theta$ and $\gamma$ for
networks for yeast, fly and human are consistent with estimates from
empirical data and our FSS analysis.

If we accept the fact that the fixed-size version of the DDDH model
in spite of its simplicity is able to reproduce certain observed
topological features of PINs, this in turn would require us to revise
the idea that the protein repertoire has evolved over millions of
years from a small set of genes to the genomes we observed today in
multi-cellular organisms which are typically composed of tens of
thousands of genes since the two are not compatible. Clearly, this is
a rather drastic measure. Rather than accept such a state of affairs,
perhaps all that the results of the fixed-size DDDH model indicate
thus far is that we should exercise caution when interpreting
minimalistic network models, as attractive as they are. Since we can
conjure up many varied and simple network models, with and without
growth, which are capable of reproducing observed features of complex
systems perhaps the only recourse when trying to pick one network
model over the other is to carefully use our knowledge of the essence
of the original real system \cite{WATTS}.


\subsection{Extensions} 
As a first step in investigating the behaviour of the DDDH model, it
seems reasonable to keep things as simple as possible, as we have done
here. We have reported on the case of fixed-size networks, and are
currently investigating how the topology is affected by varying the
relative rates of node deletion and duplication.  Moreover,
sensitivity to initial conditions is being probed further; we believe
that for the fixed-size case, the network features are independent of
initial conditions.

However, beyond the steps we have mentioned, there are obvious
extension of this model which further investigations could benefit
from including. For example, the model is based on an undirected
network -- it would be interesting to see how best to incorporate
dynamics based on directed links and what affect this would have on
the in-degree distribution and out-degree distribution. Incorporating
this feature would make the model well-suited to describing genetic
regulatory networks, for example.

Moreover, the model assumes that the rate of node deletion and
duplication are independent of one other, and independent of any
feature of the network such as the size; one could imagine the
scenario where this is not the case. Moreover, we consider single-node
deletion and single-node duplication -- an interesting variation would
be to consider multiple-node deletion or duplication, or even
duplication of whole modules (motifs) as in Ref. \cite{RAVASZ}, for
example. 

Finally, the only cause of an increase or decrease in the number of
links is either due to a node deletion or node duplication event:
links are not added or deleted through any other mechanism. The
scenario where (directed) links are stochastically added or removed
between already existing nodes would be an interesting amendment to
investigate, particularly with regards to the resulting effect on the
degree distribution and its corresponding exponent \cite{CHUNG_LU,
WANG, TADIC}.

A final example is that there is no fitness parameter in the model,
nor any rule based on selection -- our results are independent of both
of these features at the gene/protein level, and at the network level,
yet it is widely believed that both features are driving forces in the
evolution of most, if not all, biological systems. Including these
features in a meaningful way would be a highly relevant step towards
understanding some of the thornier questions in modern biology today.

\section{Conclusions}
We have introduced and discussed a minimalistic model governed by four
parameters, based on dynamic node deletion and node duplication with
heterodimerisation. The model is intended to capture some basic
features in the evolution of protein interaction networks but we
believe that it is also suited to other types of networks in light of
the suggested modifications.

Power-law degree distributions were observed for generic parameter
values, and a novel finite-size scaling effect was observed for the
case of fixed-size networks evolving through perfect duplication and
$\theta \in (0,0.5)$. 
The existence of an attractive fixed point in the average degree was
derived based on mean-field arguments, and corroborated with numerical
simulations of the first moment of the degree distribution as
described by the two-step rate equation. 
The above results were then used to to derive a relation for the
apparent exponent, $\gamma=1-2\theta$.

Our results thus far indicate consistency with empirical data. Further
investigations are required to fully explore and understand the wider
phase-space inhabited by this model, and several suggestions have been
made to this end.\\

\noindent{\textbf{{Acknowledgements}}\\ The authors would like to
thank the referees for very useful comments on a previous version of
this manuscript. The authors are grateful for helpful discussions with
Henrik J. Jensen and Simon Laird. The authors are also grateful to
Matthew Stapleton for insightful discussions on finite-size
scaling. N. Farid is thankful to EPSRC DTG for funding.


\begin{thebibliography}{55}
\bibitem{FOOTNOTE_1} We differentiate between total node removal from
the network and removal via the merging of nodes which has been
considered elsewhere, see for example \cite{SNEPPEN, DMS}.
\bibitem{FOOTNOTE_2} For an appropriately chosen initial network size
one could imagine that this regime could be used to compare the
results of perturbation effects with the results of gene knock-out
experiments.
\bibitem{FOOTNOTE_AIMS}It is worth pointing out that originally, we
set out to present the case of a growing network, evolving under
dynamic node deletion and node duplication with heterodimerisation,
and compare these results to those obtained from duplication
models. However, since we found the results for a fixed size network
particularly striking and unusual, we have restricted our results in
this paper to this special case; we defer the results for the case of
a growing network to a further publication.
\end{thebibliography}
\end{document}